\documentclass[preprint,aps,nofootinbib,11pt]{revtex4-1}
\usepackage{geometry}                % See geometry.pdf to learn the layout options. There are lots.
\usepackage{amsmath,amssymb,amsfonts,dcolumn,color,graphicx,graphics,latexsym,placeins,epsfig,tikz}
\usetikzlibrary{arrows,shapes}
\usepackage{subfigure,rotating,bm,mathrsfs}
\usepackage[title]{appendix}

\newcommand{\be}{\begin{equation}}
\newcommand{\ee}{\end{equation}}
\newcommand{\ba}{\begin{eqnarray}}
\newcommand{\ea}{\end{eqnarray}}

\usepackage[pagebackref=false, colorlinks=true]{hyperref}
\definecolor{redish}{rgb}{0.7,0.2,0.0}  % color defined in (r=red,g=green,b=blue) model
\definecolor{bluish}{rgb}{0.2,0.5,0.8}

\hypersetup{linkcolor=redish,          % color of internal links
                  citecolor=blue,        % color of links to bibliography
                  filecolor=magenta,      % color of file links
                  urlcolor=blue}          % color of the url

\begin{document}

\title{\Large Shadows in conformally related gravity theories}
\author{Kunal Pal$^{1,a}$, Kuntal Pal$^{1,b}$, Rajibul Shaikh$^{1,2,c}$ and Tapobrata Sarkar$^{1,d}$}
\address{
$^{1}$ Department of Physics, Indian Institute of Technology Kanpur, \\ Kanpur 208016, India\\
$^{2}$ Institute of Convergence Fundamental Studies \& School of Liberal Arts,\\ Seoul National 
University of Science and Technology, Seoul 01811, Korea}
\email{$^a$kunalpal@iitk.ac.in, $^b$kuntal@iitk.ac.in, $^c$rshaikh@iitk.ac.in, $^d$tapo @iitk.ac.in}

\bigskip

\begin{abstract}

Null geodesics are invariant under a conformal transformation, and thus it might seem  
natural to assume that the observables corresponding to the shadow of a space-time are also conformally invariant. 
Here, we argue instead, that since the Arnowitt-Deser-Misner mass and the active gravitational mass of an asymptotically flat 
space-time are not, in general, invariant under such conformal transformations, the shadow radius for photon motion in 
a space-time would be quantitatively different, when viewed from two different conformally related frames, 
although the expression for the shadow radius is similar. 
We then use this fact to propose a novel method 
to constrain the relevant parameters in a gravity theory conformally related to general relativity.  
As examples of our method, we constrain the parameter space in Brans-Dicke theory, 
and a class of brane-world gravity models, by using the recent observational data of M$87^{*}$ 
by the Event Horizon Telescope.

\end{abstract}
\maketitle

%%%%%%%%%%%%%%%%%%%%%%%%%%%%%%%%%%%%%%%%%%%%%%%%%%
\section{Introduction }\label{sec1}

The recent observation of the shadow of the supermassive object at the centre of our galaxy by the event horizon 
telescope (EHT) \cite{EHT1, EHT2} has paved the way to test gravitational theories in regimes of strong gravity. 
The findings of the EHT group has been used extensively to constrain and explore different aspects of 
gravity theories, from general relativity (GR) to its various alternatives.  
Most of the metrics used to model the observed shadow structure by EHT are black holes 
(both singular and regular, and  with various structures of the associated event horizons).
However, other space-times with different global topologies (like wormholes, naked singularities, 
and compact regular stars) have also been at the focus of attention, since these also 
reproduce the observed shadow structure with suitable tuning of parameters. The current literature is
abundant in such studies, and we refer the reader to \cite{RS1}-\cite{RS3}. 

Crucial in the study of shadows in a given space-time is the nature of null geodesics therein. Now, it is well 
known that null geodesics are invariant under conformal transformations. It might therefore 
be natural to assume that this property of the geodesics is inherited by the shadow of the space-time, and 
the qualitative nature of the observables (shadow radii) are (conformal) frame independent. 
In this context, we analyse the nature of the shadow radius and argue that they give 
different quantitative results  in two conformally related frames. Broadly speaking, there are two issues here, 
the first is that fact that the usual notion of mass contained in a space-time is not invariant 
under conformal transformations, and second is regarding the various types of mass contained in a spacetime in gravity theories.

As we will describe in detail, as a consequence of the first issue, if we want to take the mass to be the  same 
in two frames, as well similar values of the theory parameters in both frames, the value of the shadow radius will 
not be same in two frames. On the other hand, if we want both the shadow radius and the mass to be the same in 
two frames, the allowed values of the parameters in two frames will be naturally constrained. We use this fact 
in conjunction with the EHT data to propose a novel method 
to constrain a class of gravity theories, as we quantify with two examples. 
This is the main motivation for this work. 

Further, about the second issue mentioned above, it is well known that there are various ways to 
describe the total mass-energy contained in a gravitational system, 
and depending on the system we are considering, these definitions may or may not be equivalent (for a recent  account 
of comparison of various definitions used in the literature  see \cite{vollick}). 
We will concentrate on two such characterisation of the mass of a gravitational system, namely the 
Arnowitt-Deser-Misner (ADM) mass and the active gravitational mass (AGM). 
The key point is that though these two masses can be same in one 
of the (conformally equivalent) frames present in the theory, they may not remain the same after a
conformal transformation. For example, in the context of the scalar tensor (ST) theories considered in this paper,  
these two definitions give the same ADM mass and the AGM in the minimally coupled Einstein frame but give different values 
in the conformally related Jordan frame. And although the description of the photon motion that gives rise to 
the shadow structure of some particular metric under consideration  does not depend on the  conformal 
frame to describe the theory, there is an inherent ambiguity about what definition of the mass that 
one needs to choose, when dealing with the same system from a conformally related frame. In one  
such frame, this gives rise to different shadow structures when one considers different definition of 
the mass energy contained in the gravitational system (in our case, for ST theories, this is the Jordan frame). 
Using that particular frame, we are then able to constrain the ST parameter using EHT data.

This paper is organised as follows. 
In the first part of the next section \ref{sec2}, we describe the basic set up of the ST theory used later in the paper,  
and the transformation relations between the scalar fields of the two conformally related frames 
are given. Next, we elaborate upon the two definitions of the mass used, namely the ADM mass and the AGM. 
The corresponding formulas will be given, which will enable us to calculate the ADM mass and the AGM for a given 
space-time metric.  Then, In section \ref{sec3}, we describe the shadow radius of the Janis-Newman-Winicour (JNW) metric, 
which is the solution of the  Einstein-massless scalar field system, in both the Jordan and Einstein frames.  
Using the difference in the shadow radii in the Jordan frame, in section \ref{sec4} we constrain two different ST theories, 
namely the Brans-Dicke theory for which the coupling function is constant,   
and another ST theory which usually arises in the brane-world gravity scenario for which the coupling function 
explicitly depends on the scalar field itself. Our main findings are then discussed in the conclusion section \ref{sec5}.

\section{Set up and the formalism}
\label{sec2}

\subsection{Jordan and Einstein frames for scalar tensor theory}

We start by considering  a scalar tensor theory of gravity, where the  total 
Lagrangian  contains a non-minimally coupled scalar field,
the standard kinetic energy term and the mass term for the scalar field, as well as the action of  
some other matter fields present in the theory. It is well known that the analysis of a scalar tensor theory can be simplified 
by describing the action in terms of a conformally transformed metric of the form
${g_{\mu\nu}}=\Omega^{-2} \tilde{g}_{\mu\nu}$,
where the conformal factor $\Omega^2$ is a real, positive  definite function of the space-time coordinates $x^{\mu}$.
These metrics are commonly referred to as the Jordan frame and the Einstein frame, respectively \cite{carroll},\cite{FujiMaeda}. 
We shall below denote all the Einstein frame quantities (except the scalar field) with a ``tilde.''  
Now, it is straightforward to show that, by properly choosing the conformal factor, the non-minimally 
coupled curvature-scalar term can be recast as a standard Einstein-Hilbert action plus the action for the 
redefined scalar field. In terms of the Jordan frame scalar field $\psi$, if we write  the conformal factor as
$\Omega^2 =h (\psi)$, then  in terms of the Einstein frame scalar field $\phi$, this would be some other function say
$\Omega ^2= f(\phi)$. Here, the exact relation between the two functions will be given 
by the nature of the scalar tensor theory that we are interested in.

For a general scalar tensor theory, the Jordan frame gravitational Lagrangian is written as 
\begin{equation}\label{LJF}
\mathcal{L}=\frac{1}{16\pi }\Big[\psi R -\frac{\omega(\psi)}{\psi} \nabla_{\mu}\psi 
\nabla^{\mu}\psi -2\Lambda(\psi) \Big]~,
\end{equation}
where $R$ is the Ricci scalar and $\omega(\psi)$ and $\Lambda(\psi)$ being two functions of the scalar field $\psi$.  
The above Lagrangian can be transformed to that in the Einstein frame by a conformal transformation of the form
\begin{equation}\label{conformalmetric}
\tilde{g}_{\mu\nu}=\psi g_{\mu\nu}~.
\end{equation}
Then, by using the standard transformation relations between the Ricci scalars, it is straightforward to 
obtain the Einstein frame Lagrangian with a minimal coupling between curvature scalar and the scalar 
field (see, e.g., \cite{FujiMaeda,Clifton, quiros}). This is given by 
\begin{equation}
\mathcal{L}=\Big[\frac{1}{16\pi}\tilde{R} -\frac{1}{2} 
\tilde{\nabla}_{\mu}\phi \tilde{\nabla}^{\mu}\phi - V(\phi) \Big]~,
\label{LEF}
\end{equation}
where $V(\phi)=\frac{2\Lambda (\psi)}{\psi^{2}}$. The scalar part of the above Lagrangian in the 
Einstein frame has been  rewritten in terms of a redefined scalar $\phi$ which is related to the Jordan frame 
scalar field $\psi$ by \cite{Clifton}
\begin{equation}\label{scalars}
\frac{d \ln \psi}{d\phi}=\sqrt{\frac{16\pi}{3+2 \omega (\psi)}}~.
\end{equation}
For a given functional form of $\omega(\psi)$, integrating this relation, we obtain the explicit relation between 
the scalar fields in both the frames  \cite{Clifton}.
Note that if we consider the conformal transformation as an exact transformation of the space-time geometry, 
rather that of proper diffeomorphisms, these two scalar fields are mere rearrangements, 
but this fact will be useful for our purpose. 

\subsection{The ADM mass and the AGM in Einstein and Jordan Frames}

In GR and it's various extensions, the mass contained in an asymptotically flat space-time has 
various quantifications, which were proposed in different contexts, and all these are not equivalent 
to each other.\footnote{Though some of them are equivalent in some special circumstances. 
See the recent work of \cite{vollick} for a comparison of various such definitions.} 
Here, we will be interested specifically in two expressions, namely the ADM mass used in the context of the 
Hamiltonian formulation of gravity and the AGM contained in an arbitrary space-time. 
The expression for the ADM mass of a space-time metric with its spatial part of the form
\begin{equation}\label{spatial}
ds^2_{\Sigma}=\lambda(r) dr^2 + r^2 \chi(r) d\Omega ^2 ~,
\end{equation}
is given by Eq. 1.1.37 of \cite{PC} (see also \cite{LU})
\begin{equation}\label{ADMmass}
M_{ADM}=\lim _{r\rightarrow \infty} \frac{1}{2}\Big[-r^2 \chi^\prime (r) +r \Big(\lambda(r)-\chi(r)\Big) \Big] ~.
\end{equation}
Here, $r$ denotes a radial coordinate, and a prime denotes a derivative with respect to it. 
On the other hand, the AGM of an asymptotically flat space-time can be obtained by expanding 
the $g_{tt}$ component of a given metric in powers of  the radial coordinate $r$, and identifying the 
coefficient of $\frac{1}{r}$ in the limit $r \rightarrow \infty$, provided $\chi(r)\sim 1+\mathcal{O}(1/r)$ as 
$r \rightarrow \infty$. Hence, if $\chi(r)\sim 1+\mathcal{O}(1/r)$ as $r \rightarrow \infty$, 
and we write $g_{tt}=-\Big[1-\frac{2M_{AGM}}{r}+\mathcal{O}(1/r^2)\Big]$ then $M_{AGM}$ is the AGM of the space-time.

Since both the definitions of mass depend on the space-time metric, they naturally have different expressions 
in conformally related frames. Furthermore, their exact forms will be dependent on the conformal factor, 
which is the scalar field in this case. To obtain a relation between these masses let us now assume that the 
three dimensional metric given above represents  
the spatial part of a $3+1$ dimensional space-time metric in the Einstein frame i.e., is a solution of the Einstein 
scalar field system. Therefore, the ADM mass and the AGM in the Einstein frame are respectively 
given by $M_{ADM}^{EF}=M_{ADM}$ and $M_{AGM}^{EF}=M_{AGM}$. In the Jordan frame, 
the spatial part of the conformally transformed  metric can be written as
\begin{equation}\label{spatialJF}
ds^2_{\Sigma}=\frac{1}{\psi(r)}\Big(\lambda(r) dr^2 + r^2 \chi(r) d\Omega ^2\Big) ~.
\end{equation}
Using the formula given in Eq. (\ref{ADMmass}), we can calculate the Jordon frame ADM mass 
corresponding to this metric to be 
\begin{equation}
M^{JF}_{ADM}=\lim _{r\rightarrow \infty} \frac{1}{2}\Bigg[-r^2 \frac{d}{dr}
\Bigg(\frac{\chi(r)}{\psi(r)}\Bigg) +\frac{r}{\psi(r)} \Big(\lambda(r)-\chi(r)\Big) \Bigg] ~~.
\end{equation} 
Now, assuming that the scalar field $\psi(r)$ has a well defined limit, such that 
$\psi(r)\simeq \psi_0+\frac{\psi_1}{r}+\mathcal{O}\left(\frac{1}{r^2}\right)$ $(\psi_0\neq 0)$ 
in the limit $r\rightarrow\infty$, the above expression can be written in terms of the ADM mass 
of the Einstein frame as
\begin{equation}\label{mADMJF}
M^{JF}_{ADM}=\frac{M^{EF}_{ADM}}{\psi_{0}}-\frac{\psi_1}{2\psi_0^2} ~~.
\end{equation}
 
On the other hand, in order to find out the AGM in the Jordan frame, we first write down the space-time metric in this 
frame in the limit $r\rightarrow \infty$. This gives
\begin{equation}
ds^2=\frac{1}{\psi(r)}\Big[-\Big(1-\frac{2M_{AGM}^{EF}}{r}+\mathcal{O}\big(\frac{1}{r^2}\big)\Big)dt^2+
\lambda(r) dr^2 + r^2 \chi(r) d\Omega ^2\Big] ~,
\end{equation}
Note that we are assuming $\psi(r)\simeq \psi_0+\frac{\psi_1}{r}+\mathcal{O}\left(\frac{1}{r^2}\right)$ 
and that $\chi(r)\sim 1+\mathcal{O}(1/r)$ in the limit $r\rightarrow\infty$. We also 
assume $\lambda(r)\sim 1+\mathcal{O}(1/r)$ in this limit. Therefore, after redefining $r=\psi_0 \bar{r}$ 
and $t=\psi_0\bar{t}$ and dropping the bar at this stage, the metric in the limit $r\rightarrow\infty$ can be written as
\begin{equation}
ds^2=-\Big[1-\frac{2}{r}\big(\frac{M^{EF}_{AGM}}{\psi_{0}}+\frac{\psi_1}{2\psi_0^2}\big)+\mathcal{O}
\big(\frac{1}{r^2}\big)\Big]dt^2+\big(1+\mathcal{O}\big(\frac{1}{r}\big)\big) dr^2 + r^2 d\Omega ^2 ~.
\end{equation}
Comparing this with $g_{tt}=-\Big(1-\frac{2M_{AGM}^{GF}}{r}+\mathcal{O}\big(\frac{1}{r}\big)\Big)$, the AGM  
in the Jordan frame is given by
\begin{equation}\label{mAGMJF}
M^{JF}_{AGM}=\frac{M^{EF}_{AGM}}{\psi_{0}}+\frac{\psi_1}{2\psi_0^2}~.
\end{equation}
It is easy to see that both the expressions for the ADM mass and AGM in the Jordan frame are different 
from their Einstein frame counter parts and unless in the special case of $\psi_1=0$ two masses are 
not equal in the Jordan frame. We can also see that when $\psi_0=1$ the sum of two masses is a quantity 
invariant under conformal transformation. As we shall see this condition is indeed satisfied for the 
examples considered below.

With this formalism, we go on to explore the equivalence between the shadows in the Einstein and Jordan frames,
in terms of the ADM mass and the AGM defined above. Our strategy is to choose a convenient solution of the Einstein-scalar 
field system (i.e., a solution of scalar tensor theory in the Einstein frame), and then using the inverse relation 
of Eq. (\ref{conformalmetric}) and Eq. (\ref{scalars}), we can find the metric and the scalar field  in the Jordan 
frame, which  corresponds to a non-minimally coupled scalar tensor theory. For the Jordan frame metric, we can calculate 
the ADM mass and the AGM in terms of the mass and the scalar field. After obtaining the shadows in these two frames, 
we will be able to compare them.

%%%%%%%%%%%%%%%%%%%%%%%%%%%%

\section{Shadows in Einstein and Jordan frames}\label{sec3}

We now compare the shadow structures in the Einstein and Jordan frames. 
As we have mentioned above, the space-time metric which we take is the solution of Einstein-scalar 
field equation obtained from the Lagrangian given in Eq. (\ref{LEF}) with $V=0$. 
The solution is given by the Janis-Newman-Winicour (JNW) metric \cite{jnw}
\begin{equation}\label{JNW}
ds^2=-\Big(1-\frac{b}{r}\Big)^\gamma dt^2 + \Big(1-\frac{b}{r}\Big)^{-\gamma} dr^2 
+ \Big(1-\frac{b}{r}\Big)^{1-\gamma}r^2 d\Omega^2~.
\end{equation}
Here, the solution for the scalar field equation can be written as
\begin{equation}\label{phi}
\phi = \frac{q}{b\sqrt{4\pi}}\ln\Big({1-\frac{b}{r}}\Big)~,
\end{equation}
where $\gamma$ and $b$ are parameters appearing in the JNW metric, are related to mass $M$ and scalar charge $q$ by the relations
\begin{equation}
\gamma = \frac{2M}{b}~,~~b=2\sqrt{M^2+q^2}~.
\end{equation}
Inverting the above relations, we can write the scalar charge $q$ in terms of the mass $M$ (in the Einstein frame) and 
the parameter $\gamma$ as
\begin{equation}\label{q}
q=M\sqrt{\frac{1}{\gamma^2}-1}~.
\end{equation}
Note that the Schwarzschild solution is recovered in the limit $q\rightarrow 0$. 
Using the results from the previous section, we see that the ADM mass and the AGM of the above metric 
are the same and are given by $M_{ADM}^{EF}=M_{AGM}^{EF}=M$.

Let us now find out the photon sphere and the shadow in the Einstein frame metric given above. 
The photon sphere of a static spherically symmetric space-time is defined as orbits of the unstable 
light rays, and can be obtained by finding the effective potential encountered by the null geodesics. 
The radius of the unstable circular orbit for the Einstein frame metric (the JNW solution) 
is given by \cite{Virbhadra}
\begin{equation}
r_{PH}=\frac{b(1+2\gamma)}{2}~.
\end{equation}
Note that the photon sphere for the Einstein  frame metric in Eq. (\ref{JNW}) is defined only for 
$\frac{1}{2}< \gamma \leq 1$ which is same as the condition $0 \leq (\frac{q}{M})^2 < 3$. 
The radius of the shadow in the Einstein frame is then given by \cite{Virbhadra}
\begin{equation}\label{CIP}
b_{sh}^{EF}=\frac{1}{\gamma}\Big(1+2\gamma\Big)\left(1-\frac{2}{1+2\gamma}\right)^{(1-2\gamma)/2}M~.
\end{equation}
In order to find out the shadow in the Jordan frame, we note that, since the Jordan frame metric is 
conformally related to the Einstein frame metric, the null geodesics, photon sphere and the shadow in the 
Jordan frame will be the same as those in the Einstein frame. Therefore, the radius 
of the shadow in the Jordan frame is given by
\begin{equation}\label{shadowjordan}
b_{sh}^{JF}=b_{sh}^{EF}=\frac{1}{\gamma}\Big(1+2\gamma\Big)\left(1-\frac{2}{1+2\gamma}\right)^{(1-2\gamma)/2}M~.
\end{equation}
However, note that the mass $M$ appearing in the above expression is the ADM mass or the AGM in the Einstein frame, 
and not in the Jordan frame. As is clear from Eqs. (\ref{mADMJF}) and (\ref{mAGMJF}), the Jordan frame ADM mass or the 
AGM are different from the ones in the Einstein frame. Therefore, although in terms of the ADM mass or the AGM 
in the Einstein frame, the shadow radius has the same expression in both frames, in terms of the ADM mass or the AGM
in the respective frames, the shadow radii will be different. We can use this fact to constraint a 
scalar-tensor gravity using M87$^*$ observations by the EHT. To do so, we equate the mass predicted by the 
EHT group to either the Jordan frame ADM mass or the AGM. Note that the value of $\psi_0$ and $\psi_1$ 
will depend on the particular form of the coupling function $\omega(\psi)$. We consider two examples in the 
next section with two different choices of the coupling function.

%%%%%%%%%%%%%%%%%%%%%%%%%%%%

\section{Examples}\label{sec4}
\subsection{The Brans Dicke theory}\label{ex1}

\begin{figure}[h]
\centering
\subfigure[]{\includegraphics[scale=0.75]{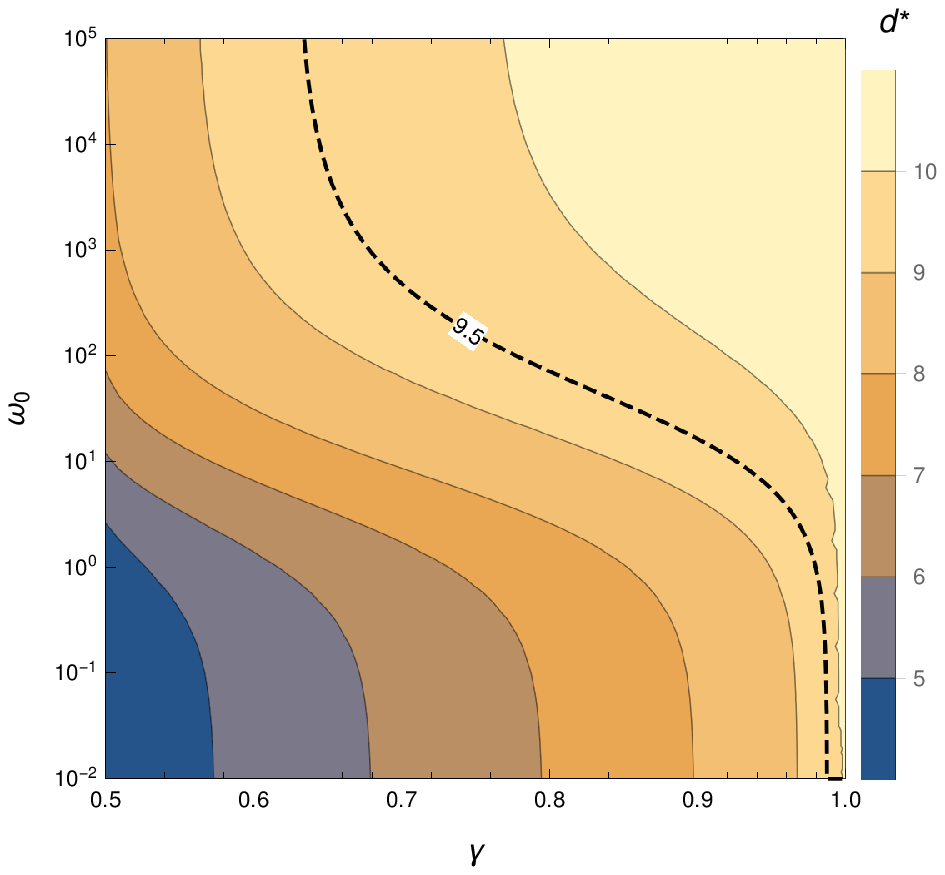}\label{fig:model1a}}
\subfigure[]{\includegraphics[scale=0.75]{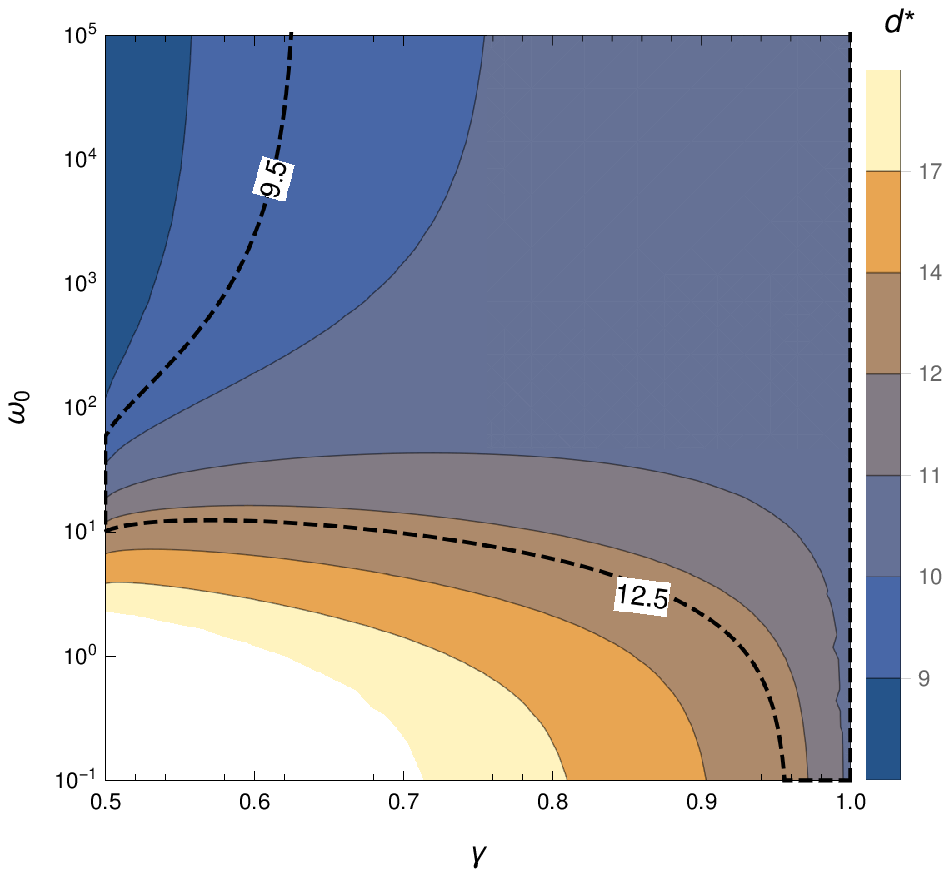}\label{fig:model1b}}
\caption{Plots showing the dimensionless diameter of the shadow when the mass of M87$^*$ is taken to be 
either (a) the ADM mass or (b) the AGM in the Jordan frame. In order that the shadow diameter is 
consistent with the M87$^*$ observation, i.e., the dimensionless diameter of the shadow is in the 
range $11.0\pm 1.5$, the allowed parameter region in (a) is that on the right side of the dashed curve and 
that in (b) is the region between the two dashed curves.}
\label{fig:model1}
\end{figure}

We now illustrate the above procedure with the Brans-Dicke (BD) theory, which is a particular example 
of the general scalar tensor theory described in section \ref{sec2}. The gravitational Lagrangian  
of the BD theory is obtained from Eq. (\ref{LJF}) by choosing $\omega=\text{constant}=\omega_{0}$ and 
$\Lambda=0$ \cite{Clifton}. These values indicate that in the Einstein frame, the BD theory describe a free 
scalar field. The relation between the scalar fields is easy to find in this case by integrating  
Eq. (\ref{scalars}). Setting the integration constant to be zero, this is given by
\begin{equation}
\ln \psi =\sqrt{\frac{16\pi}{3+2 \omega_{0}}} \phi~.
\end{equation}
Hence, using Eq. (\ref{phi}), we find that the Jordan frame scalar field is given by
\begin{equation}
\psi = \Big(1-\frac{b}{r}\Big)^{\alpha/b}~~,\text{with}~~ \alpha=\frac{2q}{\sqrt{2\omega_0+3}}~.
\end{equation}
Expanding this in the limit $r\rightarrow\infty$, we get
\begin{equation}
\psi(r)=1-\frac{\alpha}{r}+\mathcal{O}\left(\frac{1}{r^2}\right)~.
\end{equation}
Therefore, we have $\psi_0=1$ and $\psi_1=-\alpha=-\frac{2M}{\gamma}\frac{\sqrt{1-\gamma^2}}{\sqrt{2\omega_0+3}}$.
Hence, the ADM mass and the AGM in the Jordan frame respectively become (see Eqs. (\ref{mADMJF}) and (\ref{mAGMJF}))
\begin{equation}
M_{ADM}^{JF}=\left(1+\frac{1}{\gamma}\frac{\sqrt{1-\gamma^2}}{\sqrt{2\omega_0+3}}\right)M~,~~
M_{AGM}^{JF}=\left(1-\frac{1}{\gamma}\frac{\sqrt{1-\gamma^2}}{\sqrt{2\omega_0+3}}\right)M~.
\end{equation}
Now, inverting the above relations and using them in Eq. (\ref{shadowjordan}), 
we find that the Jordan frame shadow radius in terms of the Jordan frame ADM mass and the AGM are respectively given by
\begin{eqnarray}
b_{sh}^{JF}=\frac{1}{\gamma}\Big(1+2\gamma\Big)\Big(1-\frac{2}{1+2\gamma}\Big)^{(1-2\gamma)/2}
\frac{M_{ADM}^{JF}}{\left(1+\frac{1}{\gamma}\frac{\sqrt{1-\gamma^2}}{\sqrt{2\omega_0+3}}\right)}~,\nonumber\\
b_{sh}^{JF}=\frac{1}{\gamma}\Big(1+2\gamma\Big)\Big(1-\frac{2}{1+2\gamma}\Big)^{(1-2\gamma)/2}
\frac{M_{AGM}^{JF}}{\left(1-\frac{1}{\gamma}\frac{\sqrt{1-\gamma^2}}{\sqrt{2\omega_0+3}}\right)}~.
\end{eqnarray}
We now assume that the mass of M87$^*$ as predicted by EHT collaboration is either the ADM mass or the 
AGM\footnote{If we take $M$ to be equal to the average of ADM mass and AGM, then since this quantity is 
invariant under conformal transformations with $\psi_0=1$, it gives identical shadow radii in both the frames. 
However, we are not aware of a physical justification of this in the context of shadows, and 
in this paper we will not consider the possibility that mass predicted by EHT is this average value.} of the Jordan 
frame metric, and constrain the parameters $\gamma$ and $\omega_0$. To this end, we use the (dimensionless) 
diameter $d^*$ (=$\frac{2b_{sh}^{JF}}{M^{JF}_{ADM}}$ or $\frac{2b_{sh}^{JF}}{M^{JF}_{AGM}}$) 
of the shadow. $d^*$ of M87$^*$ as measured by EHT is in the range of $11\pm1.5$ \cite{bambi}. 
We use this information to constrain the parameters. The results are shown in the figure \ref{fig:model1}.
%%%%%%%%%%%%%%%%%%%%%%%%%%%%%%%%%%%%%%%%%%%%%%%%

Plots with the ADM mass and the AGM in Figs. \ref{fig:model1a} and \ref{fig:model1b} respectively, 
show qualitatively different behaviour for the  range of the BD parameter $\omega_{0}$ considered. 
In particular, in this range of $\omega_{0}$, the allowed ranges of $\gamma$ are markedly different. 
First, for both the plots, we note that in the  GR limit $\omega_{0} \rightarrow \infty$, 
the dashed line of $d^{*}=9.5$ saturates around the value $\gamma_{0}\simeq 0.63$.  
For the plot with ADM mass as the mass predicted by EHT, this is the minimum allowed value of 
$\gamma$ ($\gamma_{\min}=\gamma_{0}$) that one can get for an arbitrarily large value of $\omega_{0}$. 
However, for the plot with AGM the entire range from $0.5<\gamma<1$ is allowed, at least for 
some range of $\omega_{0}$. Also some simultaneous ranges of $\gamma$ and $\omega_{0}$ are forbidden 
in this case (the left region of $d^{*}=9.5$ and the region below the curve with $d^{*}=12.5$).

\subsection{Scalar-tensor gravity with $\omega(\psi)=-\frac{3\psi}{2(1+\psi)}$}
\label{ex2}

As a second example, we consider a theory which is more general than the BD theory, 
namely the on-brane scalar-tensor gravity developed in \cite{kanno}. In this theory, 
the function $\omega(\psi)$ in the original action in the Jordan frame given in Eq. (\ref{LJF}) 
depends on the scalar field in this frame in the following way
\begin{equation}
\omega(\psi)=-\frac{3}{2}\frac{\psi}{1+\psi}~.
\end{equation}
Here, the relation between the scalar fields in both frames can be obtained by integrating the general 
relation between them in Eq. (\ref{scalars}) and is given after a simple calculation, by
\begin{equation}\label{phi2}
\phi(\psi)=\sqrt{\frac{3}{16\pi}}\log\Big|\frac{x-1}{x+1}\Big|-\sqrt{\frac{3}{16\pi}}\log C~~, \text{where}~~~x=\sqrt{\psi+1}~,
\end{equation}
and $C$ is a integration constant. Inverting this relation we have 
\begin{equation}\label{psi2}
\psi(\phi)=\frac{4C\exp\Big(\sqrt{\frac{16\pi}{3}}\phi\Big)}{\Big(1-C\exp\Big(\sqrt{\frac{16\pi}{3}}\phi\Big)\Big)^{2}}~.
\end{equation}
The integration constant can be fixed by noting from Eq. (\ref{phi}) that,   
at $r\rightarrow \infty$, the field $\phi \rightarrow 0$, and hence the value of $\psi$ at that limit (i.e., $\psi_{0}$) 
is related to the integration constant $C$ by the relation 
\begin{equation}
\psi_{0}=\frac{4C}{\big(1-C\big)^{2}}~.
\end{equation} 
Substituting $\phi(r)$ from Eq. (\ref{phi}) and making the usual expansion 
in powers of $r^{-1}$ and finally substituting the constant $C$ by inverting the above relation we arrive at
\begin{equation}\label{psi2exp}
\psi(r)=\frac{4C\big(1-\frac{b}{r}\big)^{\frac{2q}{b\sqrt{3}}}}{\Big(1-C\big(1-\frac{b}{r}\big)
^{\frac{2q}{b\sqrt{3}}}\Big)^{2}}~~=\psi_{0}\bigg[1-\frac{2q}{\sqrt{3}r}\sqrt{1+\psi_{0}}\bigg]+\mathcal{O}\Big(\frac{1}{r^{2}}\Big)
\end{equation}

Note that $\psi_1=-\frac{2q\psi_0\sqrt{1+\psi_{0}}}{\sqrt{3}}=-\frac{2M\psi_0\sqrt{1+\psi_{0}} 
\sqrt{1-\gamma^2}}{\sqrt{3}\gamma}$. Using this expansion for the scalar field it is easy to find the 
ADM mass and the AGM for the Jordan frame of this theory by using the general formulas given in 
Eqs. (\ref{mADMJF}) and (\ref{mAGMJF}), and these are given by
\begin{equation}
M^{JF}_{ADM}=\left[1+\frac{\sqrt{1+\psi_{0}} \sqrt{1-\gamma^2}}{\sqrt{3}\gamma}\right] \frac{M}{\psi_{0}}~,~~
M^{JF}_{AGM}=\left[1-\frac{\sqrt{1+\psi_{0}} \sqrt{1-\gamma^2}}{\sqrt{3}\gamma}\right] \frac{M}{\psi_{0}}~~.
\end{equation}

\begin{figure}[h]
\centering
\includegraphics[scale=0.8]{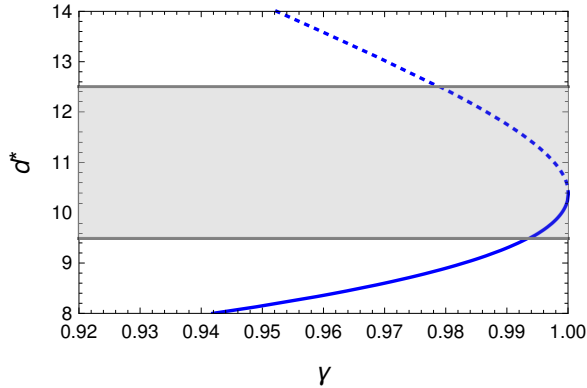}
\caption{Plot showing the dimensionless diameter of the shadow when the mass of M87$^*$ is taken to be 
either the ADM mass (solid blue) or the AGM (dashed blue) in the Jordan frame. The shaded region shows 
the allowed region bounded by the dimensionless diameters $11.0\pm 1.5$.}
\label{fig:model2}
\end{figure}

Once again, inverting these relations and substituting $M$ in Eq. (\ref{shadowjordan}), we get the shadow 
radius in terms of the Jordan frame ADM mass and the AGM for this theory to be
\begin{equation}
b_{sh}^{JF}=\frac{\Big(1+2\gamma\Big)}{\gamma}\Big(1-\frac{2}{1+2\gamma}\Big)^{(1-2\gamma)/2}
\frac{M^{JF}_{ADM}}{1+\frac{\sqrt{2(1-\gamma^2)}}{\sqrt{3}\gamma}}~,
\end{equation}
and
\begin{equation}
b_{sh}^{JF}=\frac{\Big(1+2\gamma\Big)}{\gamma}\Big(1-\frac{2}{1+2\gamma}\Big)^{(1-2\gamma)/2}
\frac{M^{JF}_{AGM}}{1-\frac{\sqrt{2(1-\gamma^2)}}{\sqrt{3}\gamma}}~,
\end{equation}
where we have taken $\psi_0=1$. Figure \ref{fig:model2} shows the dimensionless diameter 
$d^*$ (=$\frac{2b_{sh}^{JF}}{M^{JF}_{ADM}}$ or $\frac{2b_{sh}^{JF}}{M^{JF}_{AGM}}$) of the shadow. 
We find that the shadow diameter is consistent with that of the M87$^*$ observation if $\gamma$ lies 
in the ranges $0.994\leq \gamma\leq 1.0$ and $0.979\leq \gamma\leq 1.0$ when the mass of M87$^*$ is 
respectively taken to be the ADM mass and the AGM in the Jordan frame.
%%%%%%%%%%%%%%%%%%%%%%%%%%%%%%%%%%%%%%%%%%%%%%%%
%%%%%%%%%%%%%%%%%%%%%%%%%%%%%%%%%%%%%%%%%%%%%%%%%

Finally we comment upon the role of  the integration constant $\phi_{0}$ (or $\psi_{0}$). This constant 
determines the relation between the effective 
gravitational constants in the two frames. In the calculations of the weak field limits of ST theories similar 
expansions of the scalar field is performed where now the constant $\phi_{0}$ denotes the constant background 
(Minkowski) value  of the scalar field. For BD theory after such calculation is performed the effective 
gravitational constant is found to be (with $G=1$)
\begin{equation}
8\pi G_{eff}=\frac{1}{\psi_{0}}\Bigg(\frac{4+2\omega_{0}}{3+2\omega_{0}}\Bigg)
\end{equation}
Now taking the GR limit $\omega_{0} \rightarrow \infty$ we see that $G_{eff} \approx 1/\psi_{0}$. 
Hence, unless $\psi_{0}=1$ we wont get back the $G=1$  in GR. In BD theory that is why the integration 
constant is taken to be zero. (See ref \cite{quiros} for details of the calculations).
If we consider the Einstein frame gravity as GR plus matter field contribution from the scalar field, 
this is not a problem. However, if we start from Jordan frame and come back to the Einstein frame, 
then effective gravitational constant is not $G$ in this frame unless we choose the integration 
constant to be zero. For details, see \cite{SSC}.

\section{Conclusions}\label{sec5}

The fact that null geodesics remain invariant under a conformal transformation translates
to the fact that the formal expression for the shadow radius of a space-time remains invariant in conformally related frames.
What we have argued in this paper is that in spite of this formal similarity, 
the appearance of the mass in that formula however, gives rise to different shadow radii,
in the Einstein or the Jordan frames. Exploiting this transformation of mass in the two frames,
we have shown how to constrain the parameters of two 
theories of gravity, the BD theory and a class of on-brane scalar tensor gravity, using the observed data from the EHT. 
We have further elaborated our points considering two distinct definitions of mass, the ADM mass and the AGM. 
 Indeed, we find that much of the parameter space of these two
theories are ruled out, purely from observational considerations. 

Our method should be contrasted with, and is distinct from, others that appear in the literature
constraining the BD parameter, namely using Shapiro time delay \cite{Bertotti}, cosmic 
microwave background radiation \cite{Acquaviva}, \cite{Avilez}, accreting pulsars 
\cite{Psaltis}, and cosmological constraints \cite{Ooba}, \cite{Amirhashchi} (see also
\cite{Sau} which constrains the JNW solution using EHT data). Our formalism is distinct
from all these, and exploits a fundamental property regarding the difference in mass in
conformally related space-times. 

In this paper, we have performed calculations for the BD theory. However, it will be interesting to 
analyse the situation for more general coupling functions such as the one in our example in
subsection \ref{ex2}. We leave this for a future study. 

\section*{Acknowledgement}
The work of RS is partially supported by the grant from the National Research Foundation funded by the Korean government, 
Grant No. NRF-2020R1A2C1013266 (I. C.).

%%%%%%%%%%%%%%%%%%%%%%%%%%%


\begin{thebibliography}{99}

\bibitem{EHT1}
The Event Horizon Telescope Collaboration, Astrophys. J. Lett. \textbf{875}, L1 (2019).
\bibitem{EHT2}
The Event Horizon Telescope Collaboration, Astrophys. J. Lett. \textbf{875}, L6 (2019).
%\bibitem{Chandrasekhar}
%S. Chandrasekhar, \textit{The mathematical theory of black holes}, OUP, 1983.
\bibitem{RS1}
R. Shaikh, Phys. Rev. D \textbf{100}, 024028 (2019). 
%\bibitem{Newman}
%E.~T.~Newman and A.~I.~Janis,
%``Note on the Kerr spinning particle metric,''
%J. Math. Phys. \textbf{6}, 915-917 (1965).
%\bibitem{HM}
%K. Hioki and K. I. Maeda, Phys. Rev. D \textbf{80}, 024042 (2009).
\bibitem{TJ1}
T.  Johannsen, Phys. Rev. D \textbf{88} 044002 (2013). 
\bibitem{JP}
T. Johannsen and D. Psaltis, Phys. Rev. D \textbf{83}, 124015 (2011). 
\bibitem{CY}
Z. Carson, K. Yagi, Phys. Rev. D \textbf{101}, 084030 (2020).
%\bibitem{KRZ}  
%R. Konoplya, L. Rezzolla, A. Zhidenko, Phys. Rev. D \textbf{93}, 064015 (2016).
\bibitem{YZRKM}
Z. Younisi, A. Zhidenko, L. Rezzolla, R. Konoplya, Y. Mizuno, Phys. Rev. D \textbf{94}, 084025 (2016).
\bibitem{TZ}
S.  Tian, Z. Zhu,  Phys. Rev. D \textbf{100}, 064011 (2019).
\bibitem{AKVM}
A. Allahyari, M. Khodadi,  S. Vagnozzi, D. F. Mota, JCAP, \textbf{2002} (2020) 003.
\bibitem{KAVM}
M. Khodadi, A. Allahyari, S. Vagnozzi, D. F. Mota, JCAP, \textbf{2009} (2020) 026.
\bibitem{TFT}
F. Tamburini, F. Feleppa, B. Thide, arXiV : 2103.13750.
\bibitem{MT}
J. W. Moffat, V. T. Toth, Phys. Rev. D \textbf{101}, 024014 (2020).
\bibitem{JR}
S. K. Jha, A. Rahaman, Eur. Phys. J. C \textbf{81}, 345 (2021).
\bibitem{AG}
M. Afrin, S. G. Ghosh, arXiv : 2110.05258.
\bibitem{HGC}
Y. Hou, M. Guo, B. Chen, Phys. Rev. D \textbf{104}, 024001 (2021).
\bibitem{RS2}
R. Shaikh, P. Kocherlakota, R. Narayan and P. S. Joshi, Mon. Not. Roy. Astron. Soc. \textbf{482}, 52 (2019). 
\bibitem{NS1}
G. Gyulchev, P. Nedkova, T. Vetsov and S. Yazadjiev, Phys. Rev. D \textbf{100}, 024055 (2019). 
\bibitem{PTK}
V. Perlick, O. Y. Tsupko, G. S. Bisnovatyi-Kogan, Phys. Rev. D \textbf{97}, 104062 (2018). 
\bibitem{RS3}
R. Shaikh, Phys. Rev. D \textbf{98}, 024044 (2018). 
\bibitem{vollick}
D. N. Vollick, arXiv : 2101.12570.
\bibitem{carroll}
S. Carroll, \textit{Space-time and Geometry}, Pearson, 2004. 
\bibitem{FujiMaeda}
Y. Fujii, K. Maeda, \textit{The Scalar-Tensor Theory of Gravitation.} Cambridge, 2003.
\bibitem{Clifton}
T. Clifton, P. G. Ferreira, A. Padilla, C. Skordis,  Phys. Rept. \textbf{513} (2012) 1-189.
\bibitem{quiros}
I.~Quiros,
%``Selected topics in scalar–tensor theories and beyond,''
Int.\ J.\ Mod.\ Phys.\ D {\bf 28}, no. 07, 1930012 (2019).
\bibitem{PC}
P. T. Chrusciel,\textit{Lectures on Energy in General Relativity} (2010), available at \\
https://homepage.univie.ac.at/piotr.chrusciel/teaching/Energy/Energy.pdf~~.
\bibitem{LU}
J.~X.~Lu,
%``ADM masses for black strings and p-branes,''
Phys. Lett. B \textbf{313}, 29 (1993).
\bibitem{jnw}
A.I. Janis, E.T. Newman, J. Winicour, Phys. Rev. Lett. \textbf{20}, 878 (1968).
\bibitem{Virbhadra}
K. S. Virbhadra, G. F. R. Ellis, Phys. Rev. D \textbf{65}, 103004 (2002).
\bibitem{bambi}
C. Bambi, K. Freese, S. Vagnozzi, L. Visinelli, Phys. Rev. D \textbf{100}, 044057 (2019).
\bibitem{kanno}
S. Kanno and J. Soda, Phys. Rev. D {\bf 66}, 083506 (2002).
\bibitem{SSC}
A. Stabile, An. Stabile, S. Capozziello, Phys. Rev. D \textbf{88}  124011 (2013).
\bibitem{Bertotti}
B. Bertotti, L. Iess, and P. Tortora, Nature {\bf 425}, 374 (2003).
\bibitem{Acquaviva}
V.~Acquaviva, C.~Baccigalupi, S.~M.~Leach, A.~R.~Liddle and F.~Perrotta,
%``Structure formation constraints on the Jordan-Brans-Dicke theory,''
Phys. Rev. D \textbf{71}, 104025 (2005).
\bibitem{Avilez}
A. Avilez, and C. Skordis, Phys. Rev. Lett. {\bf 113}, 011101 (2014).
\bibitem{Psaltis}
D. Psaltis, Astrophys. Journal {\bf 688} 1282 (2008).
\bibitem{Ooba}
J. Ooba, K. Ichiki, T. Chiba, and N. Sugiyama, 
Prog. Theor. Exp. Phys. {\bf 2017}, 043E03, 2017.
\bibitem{Amirhashchi}
H.~Amirhashchi, and A.~K.~Yadav,
%``Constraining an exact Brans\textendash{}Dicke gravity theory with recent observations,''
Phys. Dark Univ. \textbf{30}, 100711 (2020).
\bibitem{Sau}
S. Sau, I. Banerjee, and S. Sengupta, Phys. Rev. D \textbf{102}, no.6, 064027 (2020).
\end{thebibliography}
\end{document}